\documentclass[supplement]{ptptex}

\usepackage{graphicx}
\usepackage{wrapft}

\title{
Scaling approach to glassy stationary states of
spin-glasses under chaos effects
}

\author{%       %Use \scshape  for the family name
Hajime
\textsc{Yoshino}$^{1,}$
and  Petra  E. \textsc{J{\"o}nsson}$^{2,}$
}

\inst{%         %Affiliation, neglected when [addenda] or [errata]
$^1$Department of Earth and Space Science, Faculty of Science,
Osaka University, Toyonaka, 560-0043 Osaka, Japan

$^2$ISSP, University of Tokyo, Kashiwa-no-ha 5-1-5,
Kashiwa, Chiba 277-8581, Japan
}

\abst{Dynamics of spin-glasses subjected to slow continuous changes of working
enviroment such as slow changes of temperature or interaction bonds
are studied based on scaling arguments and numerical simulations of
continuous bond changes.
Such perturbations lead to a glassy 
stationary state where the {\it age} or domain size of the system 
is pinned to macroscopic but finite values
due to competition between relaxation and chaos effects (rejuvenation). 
Flutuation-dissipation relation is also pinned 
to that of a finite age. 
The scenario explains the anomalously weak cooling rate dependece 
of spin glasses.
}

\begin{document}

\newcommand{\kb}{k_{\rm B}}
\newcommand{\tw}{t_{\rm w}}
\newcommand{\eq}[1]{Eq.~(\ref{#1})}

\maketitle

{\bf Introduction}- Dynamics of glassy systems have been 
actively studied over the past several decades. 
One general interest has been the aging effect \cite{bouetal97}, i.e.
a glassy system ages extremely slowly in a fixed working 
environment specified for example by the temperature of the heatbath.
Recently responses of such glassy systems
to various kinds of perturbations have attracted much  attention
including restart of aging or rejuvenation of spin-glasses 
after sudden changes of the temperature of the heatbath 
\cite{jonetal98,jonetal2003}, 
responses of soft glassy systems 
to shear\cite{SLHC97}.

In the present work we study relaxation of spin-glasses subjected to
slow changes of the working environment, namely slow changes of 
temperature or interaction bonds. This study is motivated  by the recent
experimental observations that spin-glasses exhibit surprisingly 
weak dependence on the cooling rates \cite{jonetal98,jonetal2003}. 
We propose a scaling ansatz
for the effective age of the system under continuous changes of the
working environment 
assuming that the competition between aging and rejuvenation due 
to the chaos effects  \cite{droplet}
leads to a stationary state. 
We performed Monte Carlo (MC) simulations of a
4-dimensional Edwards-Anderson (EA) model subjected to continuous changes
of bonds and found good agreement with the scenario.

{\bf Model}- Specifically we consider the Edwards-Anderson (EA) spin-glass
model with Ising spins $S_{i}$ ($i=1 \ldots N$) on a $d$-dimensional 
(hyper)-cubic lattice with $N=L^{d}$ lattice points. The Hamiltonian 
is given by $H=-\sum_{<i,j>}J_{ij} S_{i}S_{j}-h\sum_{i}S_{i}$.
We use the $\pm J$ model
where the interaction bonds $J_{ij}$ between nearest-neighbours 
$\langle i,j \rangle$ take random  $\pm J$ values.
The temperature is measured as $\kb T/J$ with
the Boltzmann's constant set as $\kb =1$. The 
magnetic field $h$ is measured as $h/J$. The relaxational dynamics is
modeled by the usual heatbath single spin flip MC algorithm.

A continuous bond changes is realized by flipping the sign 
of a fraction $p \ll 1$ of the bonds chosen randomly 
in 1 Monte Calro step (MCS).
This amounts to flip a fraction $p_{\rm eff}(t)=[1-(1-2p)^{t}]/2$
of the original set of bonds $J_{ij}(0)$  to
create the set of bonds $J_{ij}(t)$ at $t$ (MCS).
For $pt \ll 1$ $p_{\rm eff}(t)=pt$ and 
$p_{\rm eff}(t) \to 1/2$ for $pt \gg 1$.

A continuous temperature changes is specified by a 
cooling or heating rate $v_{T}=\delta T/\delta t$ 
where $\delta T$ is the change of the temperature within an
interval of time $\delta T$.
Thus the resultant change of the temperature after a time 
$t$ is $\Delta T(t)=v_{T}t$.
Unfortunately the temperature-chaos effect is much weaker than
the bond chaos effect \cite{chaos} in the present model
and we limit ourselves to the bond perturbations in the simulations.

{\bf Scaling arguments}- 
Let us consider a spin-glass whose spins are evolving in a fixed 
working environment 
$W=({\cal J}, T)$ specified by a set of bonds ${\cal J}=\{J_{ij}\}$
and the heatbath temperature $T$. If the initial spin configuration
is far from the equilibrium state associated with $W$, the system
is equilibrated only up to a certain length scale $L_{T}(t)$ after a time $t$.
The domain size $L_{T}(t)$
grows slowly with time $t$ by thermally activated droplets excitations
as $L_{T}(t)=L_{0}[(T/J)\log(t/\tau_{c}(T))]^{1/\psi}$ where 
$L_{0}$ is a unit of length scale,  $\tau_{c}(T)$
is the attempt time of the activated dynamics and $\psi > 0$ is the exponent
of the free-energy barriers (See \cite{yoshuktak2002} for the details).

Now we perturb the system by slowly changing $W$ either by continuous
temperature changes characterized by $\Delta T(t)=v_{T}t$ 
or continuous bond changes characterized by 
$\Delta J(t)=J\sqrt{p_{\rm eff}(t)}$ \cite{chaos}.
Within an interval of time $\tau$, the temporal working environment 
$W(t)$ changes from  $W(0)$ to $W(\tau)$.
Naturally the overlap length  \cite{droplet} between the target
equilibrium states associated with the two environments can be introduced as
$L_{\Delta J(\tau)} \sim L_{0}(\Delta J(\tau)/J)^{-1/\zeta}$ 
and $L_{\Delta T(\tau)} \sim L_{0}(\Delta T(\tau)/T_{\rm g})^{-1/\zeta}$.
The target equilibrium states are completely different
at length scales much larger than the overlap lengths 
by the chaos effects\cite{droplet}.
During the continuous perturbations,
the relaxational process which tends to grow the domains of 
$W(t)$ and rejuvenation due to 
the chaos effects must be competing with each other. Naturally we
expect that the system attains a certain stationary state characterized
by a certain fixed size of the domains $L_{T}(\tau^*)$ with respect 
to the temporal working environment $W(t)$. In other words
the {\it age} of the system is pinned to a certain value $\tau^{*}$.
As proposed in Ref \cite{jonetal2003,yos2003}, such a length scale 
should play the role of a sort of renormalized overlap length 
which parameterize the heating/cooling rate effects 
in temperature-cycling experiments.
Simplest possible scaling form for the stationary domain size is
\begin{equation}
L_{T}(\tau^{*}) \sim L_{0} (\Delta J(\tau^{*})/J)^{-1/\zeta} \qquad 
L_{T}(\tau^{*}) \sim L_{0} (\Delta T(\tau^{*})/T_{\rm g})^{-1/\zeta}
\label{eq-trelax}
\end{equation}
It can be seen that by choosing slower 
perturbations $p \to 0$ or $v_{T} \to 0$, 
the pinned age increases $\tau^{*} \to \infty$ and the
stationary domain size increases $L_{T}(\tau^*) \to \infty$.

It can be also seen that 
in the limit of slow perturbations $p \to 0$ or $v_{T} \to 0$ 
actual change of the working environment
becomes infinitesimal $\Delta J(\tau^*) \to 0$ or $\Delta T(\tau^*) \to 0$.
The later feature is quite unusual compared with analogues of
more conventional cases such as systems under shear with shear rate
$\dot{\gamma}$. There one typically finds that
relaxation times scales as $\tau^{*} \propto \dot{\gamma}^{-\nu}$
with $\nu \leq 1$ so that actual deformation of the system 
$\gamma \propto \dot{\gamma}^{1-\nu}$ within the relaxation time $\tau^{*}$
can {\it diverge} as $\dot{\gamma} \to 0$.

During a continuous change of temperature the growth law $L_{T}(t)$
would also change. However it is
asymptotically irrelevant in the above scenario  since
$L_{T+\Delta T}(t)/L_{T}(t)=(T+\Delta T/T)^{1/\psi} = O(1)$ .
Thus for a given cooling/heating rate $v_{T}$ the relaxation time $\tau^{*}$ 
is defined at every local temperature $T$. The effective age 
$\tau^{*}$ of a system cooled down to a certain target temperature $T_{m}$ 
will be determined only by local cooling rate in the vicinity of $T_{m}$
and the cooling rates used at higher temperatures are irrelevant.
This feature observed experimentally in spin-glasses \cite{jonetal98}
would hardly be present in systems with activated dynamics but 
without the chaos effect.

\begin{figure}
\begin{center}
\resizebox{6.5cm}{5cm}{
\includegraphics{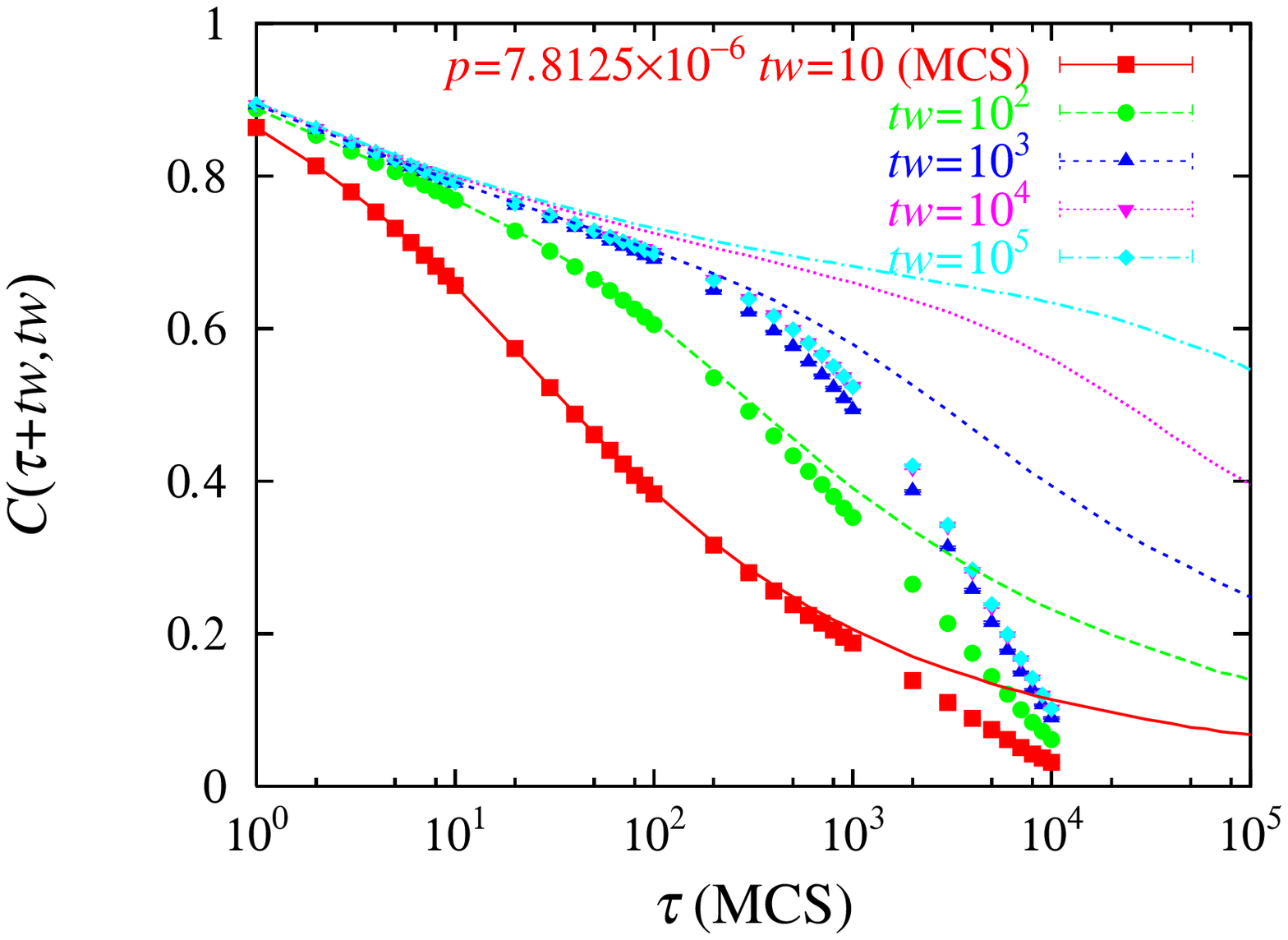}}
\resizebox{6.5cm}{5cm}{
\includegraphics{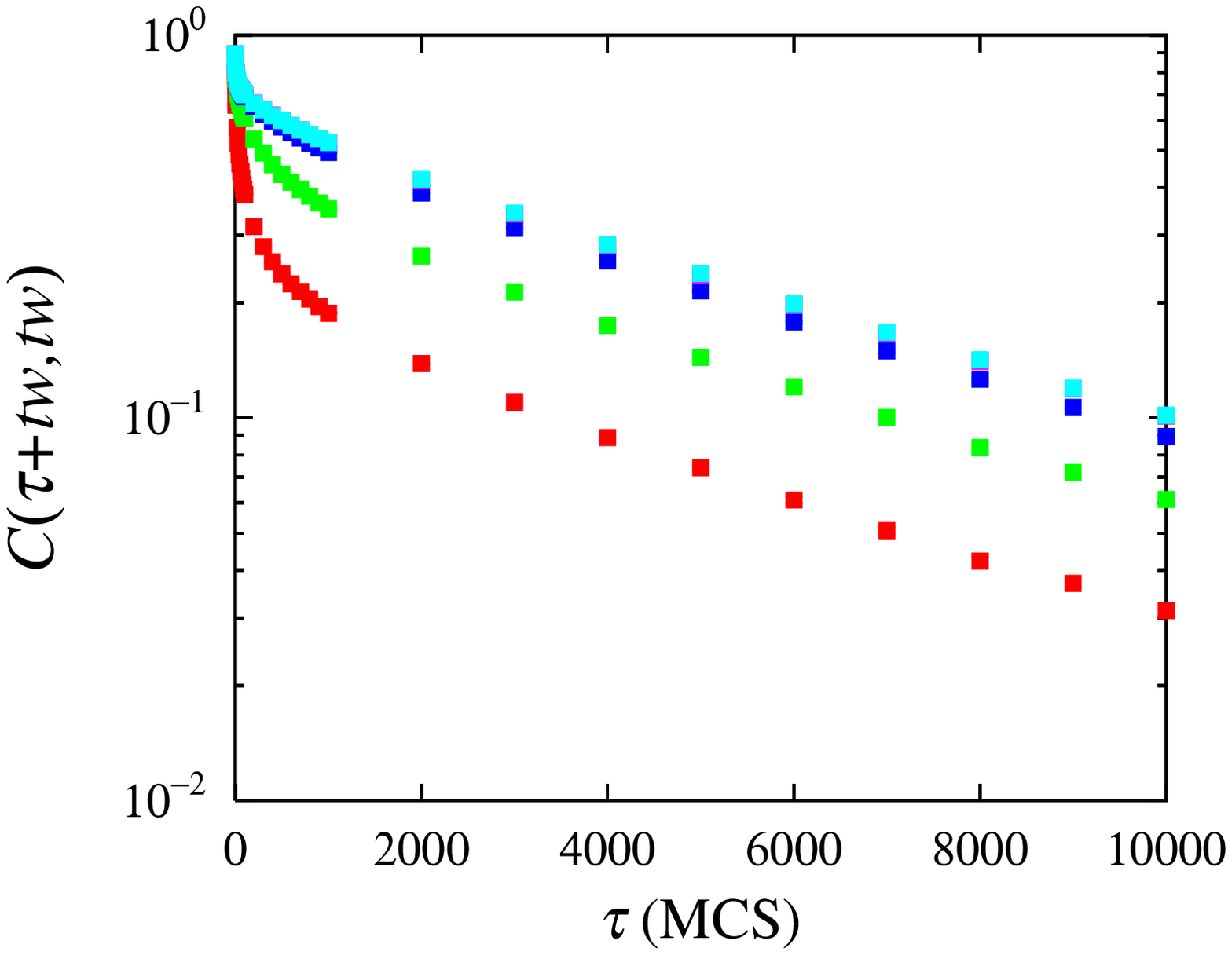}}
\end{center}
\caption{Spin autocorrelation function
$C(\tau+\tw,\tw)=(1/N)\sum_{i=1}^{N} \overline{\langle S_{i}(\tau+\tw)S_{i}(\tw) \rangle}$ under continuous bond perturbation
of strength $p=7.8125 \times10^{-6}$ at $T/J=1.2$.
Here the averages over different realizations of thermal noises
and sequences of bonds are denoted as $\langle \cdots \rangle$ and
$\overline{\cdots}$ respectively.
The lines in the left figure are reference data of standard 
isothermal aging.\cite{yoshuktak2002} 
The system exhibits interrupted aging: waiting time $\tw$ dependence 
saturates. The tail parts of the autocorrelation functions are
fitted to a simple exponential form $A \exp(-t/\tau^{*}(p))$ 
by which the characteristic
relaxation time $\tau^{*}(p)$ is extracted. In this example
$\tau^{*}=6013$ (MCS) is obtained.}
\label{fig-interruption}
\end{figure}
\begin{figure}
\begin{center}
\resizebox{6.5cm}{5cm}{
\includegraphics{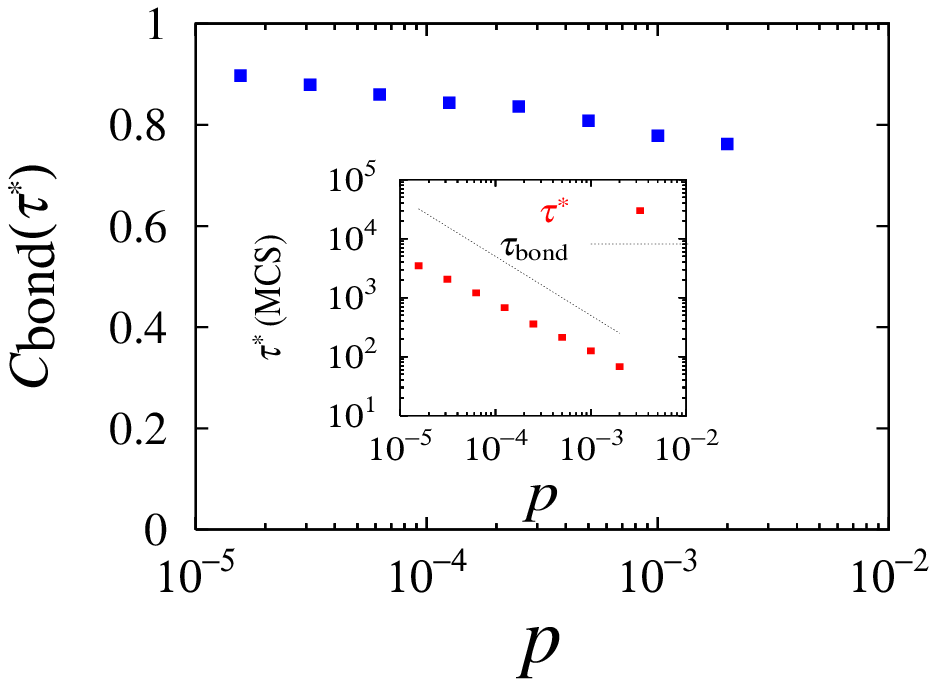}}
\resizebox{6.5cm}{5cm}{
\includegraphics{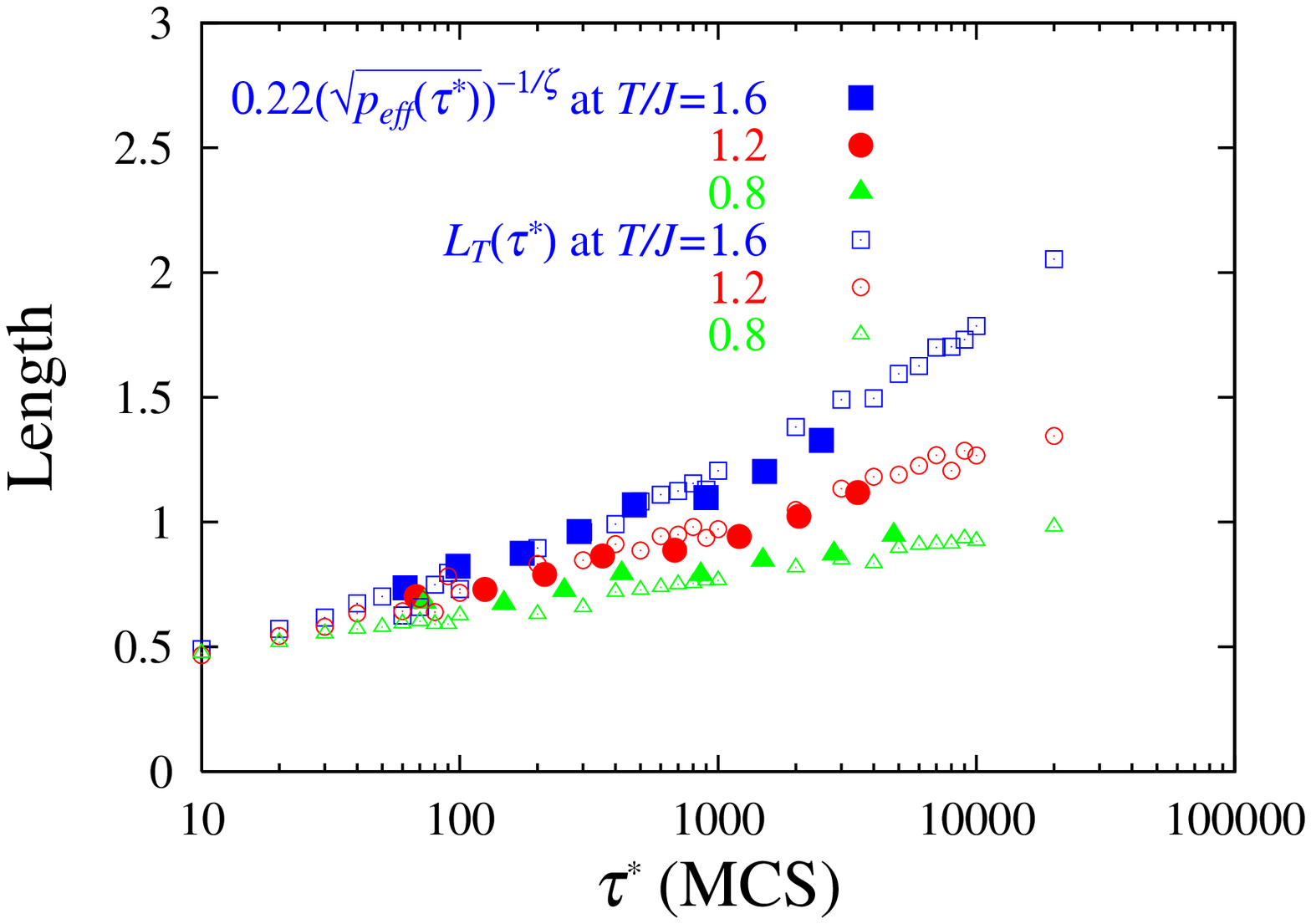}}
\end{center}
\caption{The correlation $C_{\rm bond}(\tau^{*})$ 
between the set of bonds at two times separated by
the relaxation time $\tau^*(p)$ is shown in the left figure.
It is defined as
$C_{\rm bond}(t) \equiv
\overline{J_{ij}(0)J_{ij}(t)}/J^{2}=\exp(-t/\tau_{\rm bond}(p))$ with 
the decorrelation time of the bonds 
given as $\tau_{\rm bond}(p)=-1/\log(1-2p)\simeq 1/p$.
As shown in the inset the relaxation time of the spins 
 $\tau^{*}(p)$ grows with decreasing $p$ but less rapidly than
the decorrelation time of the bonds $\tau_{\rm bond}(p)$.
In the right figure the scaling ansatz Eq. \ref{eq-trelax} is tested.
The open symbols are the reference data of $L_{T}(\tau)$ 
(from Ref \cite{yoshuktak2002}). The filled symbols
are the overlap length $a(\Delta J(\tau)/J)^{-1/\zeta}$ 
with $\Delta J(t)=J \sqrt{p_{\rm eff}(t)}$ at 
the relaxation time $t=\tau^{*}(p)$ with
$p=0.002\times 2^{-n}$ with $n=0,\ldots,7$ (from left to right)
obtained at $T/J=0.8,1.2$ and $1.6$.
Here $\zeta=0.9$ as obtained in Ref. \cite{jonetal2003}
(see also \cite{chaos}) and chosen the prefactor $a=0.22$.}
\label{fig-trelax}
\end{figure}

{\bf Numerical analysis}-
We performed MC simulations of the dynamics of the 4-dimensional 
EA model ($T_{\rm g}\simeq 2.0J$) under continuous bond changes
starting from random initial configurations.
In Fig. \ref{fig-interruption} we show the data of the spin
autocorrelation function $C(\tau+\tw,\tw)$. Apparently the
system becomes stationary within a finite time scale $\tau^{*}$
due to the continuous bond changes. The relaxation time $\tau^{*}$ 
extracted from the data are examined as shown in Fig. \ref{fig-trelax}.
It appears that the actual change of the bonds 
within the relaxation time $\tau^{*}$ of the bonds
becomes negligible as the intensity of the driving $p$ becomes smaller.
As shown in the right figure of  Fig. \ref{fig-trelax}, the relaxation
time $\tau^{*}$ follows the expected scaling \eq{eq-trelax}.

\begin{wrapfigure}{r}{\halftext}
\begin{center}
\resizebox{7cm}{5cm}{
\includegraphics{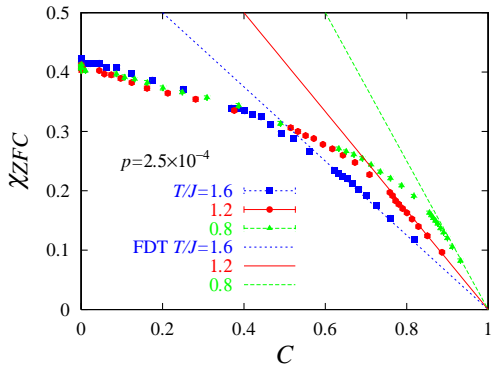}}
\end{center}
\caption{The ZFC susceptibility
$\chi_{\rm ZFC}(\tau+\tw,\tw)$ vs. the spin autocorrelation function 
$C(\tau+\tw,\tw)$ with $\tw \gg \tau^{*}$. In this example
$p=2.5 \times 10^{-4}$.
$\chi_{\rm ZFC}$ is measured as usual \cite{yoshuktak2002}
with $h/J=0.1$ but under continuous bond changes.
The lines represents FDT
$\chi_{\rm ZFC}=(1-C)/T$ associated with the heatbath temperatures $T$.}
\label{fig-ck}
\end{wrapfigure}

Finally let us examine the fluctuation dissipation relation (FDR)
as proposed by the dynamical mean field theory (DMFT) \cite{CK93}.
As shown in Fig. \ref{fig-ck} the fluctuation dissipation theorem
is satisfied at short time scales and violated at larger time scales in a 
non-trivial manner as predicted by the DMFT.
Thus the stationary state can be considered as a glassy state.
The result is very similar to that of the same model 
during isothermal aging \cite{yoshuktak2002}
and driven dynamics induced by asymmetric coupling \cite{B2003}.
It must be noted however that the FDR  systematically depends 
on $p$ as those of isothermal
aging systematically depends on $\tw$
so that the FDR in the asymptotic limit
$p \to 0$ may be different from the prediction of the DMFT
\cite{yoshuktak2002}.

{\bf Discussion-} To conclude we studied glassy stationary dynamics of
spin-glasses under continuous temperature or bond changes.
Let us consider a typical cooling rate  $v_{T}=O(10^{-1})$(K/sec) 
used in `` qunech'' experiments, which is actually
as slow as $v_{T}=10^{-15} J/{\rm MCS}$ 
since $T_{\rm g}=O(10)$ K 
and the microscopic time scale is $\tau_{0}=10^{-13}$ (sec).
The scaling ansatz \eq{eq-trelax} implies 
it yields $\tau^{*}$ of only order $1-10$ (sec)
at a target temperature $T_{\rm m}=0.7T_{\rm g}$ in the case of a canonical
spin-glass AgMn sample \cite{jonetal2003},
which allows one to observe ``clean'' aging at $T_{\rm m}$.
Indeed the effective age of the system after such slow cooling 
is found to be only $O(10)$ (sec) experimentally \cite{jonetal2003}. 

%\begin{thebibliography}{99}
%%%%%%%%%%%%%%%%%%%%%%%%%%%%%%%%%%%%%%%%%%%%%%%%%%%%%%%%%%%%%
% Some macros are available for the bibliography:
%  o for general use
%    \JL : general journals                 \andvol : Vol (Year) Page
%  o for individual journal 
%    \AJ   : Astrophys. J.           \NC         : Nuovo Cim.
%    \ANN  : Ann. of Phys.           \NPA, \NPB  : Nucl. Phys. [A,B]
%    \CMP  : Commun. Math. Phys.     \PLA, \PLB  : Phys. Lett. [A,B]
%    \IJMP : Int. J. Mod. Phys.      \PRA - \PRE : Phys. Rev. [A-E]     
%    \JHEP : J. High Energy Phys.    \PRL        : Phys. Rev. Lett.
%    \JMP  : J. Math. Phys.          \PRP        : Phys. Rep.
%    \JP   : J. of Phys.             \PTP        : Prog. Theor. Phys.     
%    \JPSJ : J. Phys. Soc. Jpn.      \PTPS       : Prog. Theor. Phys. Suppl.
% Usage:
%  \PRD{45,1990,345}          ==> Phys.~Rev.\ \textbf{D45} (1990), 345
%  \JL{Nature,418,2002,123}   ==> Nature \textbf{418} (2002), 123
%  \andvol{B123,1995,1020}    ==> \textbf{B123} (1995), 1020
%%%%%%%%%%%%%%%%%%%%%%%%%%%%%%%%%%%%%%%%%%%%%%%%%%%%%%%%%%%%%
  
%\bibitem{}

\end{document}